\documentclass{PoS}

\usepackage[authoryear]{natbib} 
\bibpunct{(}{)}{;}{a}{}{,}      

\title{Recent extensions to GALPROP}

\ShortTitle{Extensions to GALPROP}

\author{\speaker{A. W. Strong}\\
        Max-Planck-Institut f\"ur extraterrestrische Physik\\
        E-mail: \email{aws@mpe.mpg.de}}

\abstract{Some recent extensions to the GALPROP cosmic-ray propagation package  are described.
The enhancements include: an accurate solution option, improved convection formulation, alternative spatial boundary conditions, polarized synchrotron emission, 
new magnetic field models, updated gamma-ray production cross-sections, free-free radio emission and absorption, primary positrons,
additional injection spectral breaks,  deuterium production by pp fusion, 
hadronic energy losses,
improved HEALPix skymap format, compatibility with latest HEALPix release,
and various bug fixes.
The Explanatory Supplement has been extensively updated, including details of these enhancements.
A compatible plot package GALPLOT for GALPROP output is also provided, as well as other related software.

}

\FullConference{The 34th International Cosmic Ray Conference,\\
		30 July- 6 August, 2015\\
		The Hague, The Netherlands}

\def\gray{$\gamma$-ray\ }
\def\grays{$\gamma$-rays\ }

\begin{document}
\section{Introduction}
GALPROP is a software package for numerical Galactic cosmic-ray propagation and related emission processes. Its origins go back more than 20 years
\cite{ICRC1993,ICRC1995,SY1996}.
The first mature version of GALPROP was presented in   \cite{1998ApJ...493..694M,1998ApJ...509..212S}.
and it has been developed further in the following years.
It is described  in a wider context in \cite{2007ARNPS..57..285S}.
A recent related review is \cite{AnnRev2015}.

The extensions described here were made 
based on the  C++ version (v54) publicly released four years ago in 2011\footnote{http://galprop.stanford.edu. Last update 2011.}.
These developments were prompted by new requirements and features which appeared desirable, including suggestions by users.
The extended version is available\footnote{https://sourceforge.net/projects/galprop};
  it has been downloaded over 400 times since its initial release in 2013,  showing the community interest.
It is an ongoing development, with regular updates.
GALPROP  is maintained under the configuration control system {\it subversion (svn)}.
I briefly describe each of the new features, grouped according to topic,  with desiderata for future developments.
Details can be found  in the Explanatory Supplement document included in the package.

\section{Physical processes}
\subsection{Hadronic production cross-sections}
There has been progress in hadronic \gray production physics
\cite{2013arXiv1307.0497D,2013arXiv1303.6482D,2013EPJWC..5202001O} 
 and this is reflected in new options including combinations appropriate to low and high energy protons\footnote{Thanks to Chuck Dermer, Michael Kachelriess and Sergey Ostapchenko for providing the relevant code, and much help on this topic.}.
A correction for nuclei heavier than Helium in both CR and interstellar gas is provided.
Corresponding options for secondary positrons and electrons are foreseen in future.

\subsection{Hadronic energy losses}
Energy loss of nuclei by hadronic processes (pion production) was not included previously, only ionization and Coulomb losses were present, so that energy conservation was not respected
 since there is hadronic production of \grays and leptons.
Hadronic losses are comparable to  ionization and Coulomb losses around 1 GeV.
This is a potentially important process especially in regions of high gas density.
It has been included in a simple treatment based on
\cite{2002cra..book.....S,2015ApJ...802..114K}
 for protons and Helium.
This is a complex problem however, and further developments are desirable.
In particular how to treat losses in the context of nuclear fragmentation, beyond the approximation of conserved energy per nucleon, is difficult.
 
\subsection{Deuterium production by pp fusion}
At energies up to 1 GeV, deuterium production by pp fusion is significant and has been included\footnote{Thanks to Nicolas Picot-Clemente for suggesting this topic, and help in its implementation.} using the formulation in \cite{2012A&A...539A..88C}. See also Picot-Clemente et al.  this conference, paper PoS(ICRC2015)555.

\subsection{Synchrotron radiation}
The treatment of synchrotron radiation from electrons and positrons has been updated to include polarization using the regular component of the magnetic field.
For a  detailed description and results see \cite{2011A&A...534A..54S,2013MNRAS.436.2127O}.
GALPROP models of the synchrotron spectral index have been used in the  analysis of Galactic microwave emission observed by {\it Planck} \cite{2015arXiv150606660P}.

\subsection{Free-free emission and absorption}
Although not a CR-related process, radio surveys include free-free emission from thermal electrons, so skymaps are now produced. These are relevant at frequencies above 1 GHz, and dominated for WMAP and Planck.
The model is based on the NE2001 thermal electron model.
Free-free absorption is important at radio frequencies below 100 MHz, and this can be applied to both synchrotron and free-free emission. Again NE2001 is used. See \cite{2013MNRAS.436.2127O} for details and results.
Developments beyond NE2001 would be desirable in future since this has known problems.
\section{Cosmic-ray propagation}
\subsection{Accurate solution options}
The original numerical scheme for solving the CR propagation equations uses a Crank-Nicolson scheme with operator splitting, and a procedure to accelerate the solution by varying the time step from large to small values. Both the operator splitting and time-reduction schemes have the effect that the true solution to the steady-state equation is not obtained (see Explanatory Supplement and also \cite{2015APh....70...39K}). This can be checked by detailed diagnostics of the solution, provided as a GALPROP  option. In order to obtain accurate solutions, an {\it explicit} scheme is now available, which uses small enough time-steps that none of the mentioned approximations are required, and demonstrably converges to the steady-state solution. Detailed analysis  is in the Explanatory  Supplement. The disadvantage of this method is the long computational time on account of the small time steps.
This is clearly a brute-force approach, and the application of modern numerical methods, such as in PICARD \cite{2014APh....55...37K,2015APh....70...39K}, which can solve the equations fast and  accurately,  would be preferable in future.

\subsection{Anisotropic diffusion}
Spatial diffusion was assumed to be isotropic, but theory predicts that it is anisotropic, being slower perpendicular to the regular field, hence slower in the $z$-direction.
An option for separate diffusion coefficients in the Galactic plane and perpendicular to it has therefore been implemented.

In future, a more specific relation to the regular B-field would be desirable.
 
\subsection{Convective transport}
Convection (Galactic wind) was originally represented as a velocity increasing linearly from the Galactic plane, which is rather unphysical. 
A more plausible form has been introduced, which still respects the condition of zero velocity at $z=0$ (since the sign must reverse there) and which smoothly increases to a constant value beyond some $z$.
The form (using a {\it tanh} function) is controlled by two parameters for the strength and $z$-dependence of the wind.
In future a physical description of the wind would be desirable.

\subsection{Boundary conditions}
The spatial boundary condition imposed has been zero CR density at all boundaries in 2D ($R,z$) or 3D ($x,y,z$). This is clearly not physical since even with free-escape, the density is non-zero there.
The condition was imposed by simply forcing a zero value after each time-step, rather than acually solving with this condition.
As an alternative, the boundary condition is not imposed, so particles flow out but not back in at the boundary. The density at the boundary is non-zero.
A more physical approach actually handling the physics of the boundary, would be desirable in future.

\subsection{Primary positrons}
There is good experimental evidence for a primary component of positrons (PAMELA, AMS-02), in addition to secondaries from hadronic interactions.
Previously only secondary positrons were included.
Primary positrons can now be included, with an injection spectrum and spatial source distribution independent of primary electrons.
Note that primary positrons also contribute to \grays via inverse-Compton and bremsstrahlung, and to synchrotron emission, so are important to take into account.

\subsection{Injection spectrum}
Additional spectral breaks in the injection spectra of nuclei, electrons and positrons are now available, for more flexibility in fitting observations.

\section{Galactic structure}
\subsection{Magnetic fields}
Additional models of the Galactic magnetic field, including the regular field described by \cite{2012ApJ...757...14J,2012ApJ...761L..11J}, are now included.

\section{Format and examples}
\subsection{HEALPix}
Skymap output (\grays, radio) is provided in HEALPix \cite{2005ApJ...622..759G} as the format of choice for full-sky coverage with uniform pixel size. 
New versions of HEALPix have become available\footnote{http://healpix.jpl.nasa.gov}, and the GALPROP code has been adapted to enable their use.
The format has been adjusted so that the energy (or frequency for radio) is stored as FITS columns rather than the original format of vectors within pixels.
This is compatible with the Aladin visualization package\footnote{http://aladin.u-strasbg.fr}, and also more convenient for reading by user software.

\subsection{Examples and reference output}
A variety of example parameter files (galdef files) are provided to illustrate the many features including the new ones.
As an installation check, a full set of output files is also provided for one sample case.

\subsection{Bug fixes}
Various (minor) bug fixes to the original version  were made during the development.

\section{Related software}
A plotting package GALPLOT, for cosmic rays, \grays and radio, compatible with GALPROP output, is now available\footnote{https://sourceforge.net/projects/galplot}. 
It is compatible with the cosmic-ray database CRDB \cite{2014A&A...569A..32M}, and can handle Fermi-LAT and other \gray data and various synchrotron surveys.
GALPLOT can also be used for Galactic source population synthesis, as described in section 6 of \cite{2015arXiv150102003T}.

Separate routines for computing synchrotron\footnote{https://sourceforge.net/projects/galpropsynchrotron},
anisotropic inverse Compton scattering, and models for inverse-Compton emission from solar and stellar heliospheres  are also available\footnote{https://sourceforge.net/projects/stellarics}.

\section{Outlook}
Other projects complementary to GALPROP  are in progress.
DRAGON \cite{2015PhRvD..91h3012G} includes various new features.
PICARD \cite{2014APh....55...37K,2015APh....70...39K}  allows full 3D models with high resolution  using advanced numerical methods.
The semi-analytical USINE \cite{2010A&A...516A..67M} package is becoming public (see Maurin, this conference, paper ID 296).
These activities will stimulate further GALPROP developments.

\bibliographystyle{JHEP}
\bibliography{sourcepop,luminosity,strong,CRemiss,andy.bib,galprop_icrc2015} 

\providecommand{\href}[2]{#2}\begingroup\raggedright\begin{thebibliography}{10}

\bibitem{ICRC1993}
A.~W. {Strong} and G.~{Youssefi}, {\it {Spectrum of electrons with
  inhomogeneous energy losses}},  {\em Proc. 23rd ICRC} {\bf 2} (1993)
  124--127,
  [\href{http://arxiv.org/abs/http://www.mpe.mpg.de/$\sim$aws/publications/cal%
garyprop.ps}{{\tt
  http://www.mpe.mpg.de/$\sim$aws/publications/calgaryprop.ps}}].

\bibitem{ICRC1995}
A.~W. {Strong} and G.~{Youssefi}, {\it {Propagation models for cosmic-ray
  nucleons and electrons and predictions of the Galactic gamma-ray spectrum}},
  {\em Proc. 24th ICRC} {\bf 3} (1995) 48--51,
  [\href{http://arxiv.org/abs/http://www.mpe.mpg.de/$\sim$aws/publications/str%
ong\_youssefi\_icrc\_24.ps}{{\tt
  http://www.mpe.mpg.de/$\sim$aws/publications/strong\_youssefi\_icrc\_24.ps}}%
].

\bibitem{SY1996}
A.~W. {Strong} and G.~{Youssefi}, {\it {Cosmic-ray propagation: self-consistent
  models for nuclei, electrons and gamma rays}},
  \href{http://arxiv.org/abs/http://www.mpe.mpg.de/$\sim$aws/publications/prop%
agate\_paper1.ps}{{\tt
  http://www.mpe.mpg.de/$\sim$aws/publications/propagate\_paper1.ps}}.

\bibitem{1998ApJ...493..694M}
I.~V. {Moskalenko} and A.~W. {Strong}, {\it {Production and Propagation of
  Cosmic-Ray Positrons and Electrons}},  {\em \apj} {\bf 493} (Jan., 1998)
  694--+.

\bibitem{1998ApJ...509..212S}
A.~W. {Strong} and I.~V. {Moskalenko}, {\it {Propagation of Cosmic-Ray Nucleons
  in the Galaxy}},  {\em \apj} {\bf 509} (Dec., 1998) 212--228.

\bibitem{2007ARNPS..57..285S}
A.~W. {Strong}, I.~V. {Moskalenko}, and V.~S. {Ptuskin}, {\it {Cosmic-Ray
  Propagation and Interactions in the Galaxy}},  {\em Annual Review of Nuclear
  and Particle Science} {\bf 57} (2007) 285--327,
  [\href{http://arxiv.org/abs/astro-ph/}{{\tt astro-ph/}}].

\bibitem{AnnRev2015}
I.~A. {Grenier}, J.~H. {Black}, and A.~W. {Strong}, {\it {The Nine Lives of
  Cosmic Rays}},  {\em \\ Ann. Rev. Astron. Astrophys.} {\bf 53} (2015)
  199--246,
  [\href{http://arxiv.org/abs/doi:10.1146/annurev-astro-082214-122457}{{\tt
  doi:10.1146/annurev-astro-082214-122457}}].

\bibitem{2013arXiv1307.0497D}
C.~D. {Dermer}, A.~W. {Strong}, E.~{Orlando}, L.~{Tibaldo}, and {for the Fermi
  Collaboration}, {\it {Determining the Spectrum of Cosmic Rays in Interstellar
  Space from the Diffuse Galactic Gamma-Ray Emissivity}},  {\em ArXiv e-prints}
  (July, 2013) [\href{http://arxiv.org/abs/1307.0497}{{\tt arXiv:1307.0497}}].

\bibitem{2013arXiv1303.6482D}
C.~D. {Dermer}, J.~D. {Finke}, R.~J. {Murphy}, A.~W. {Strong}, F.~{Loparco},
  M.~N. {Mazziotta}, E.~{Orlando}, T.~{Kamae}, L.~{Tibaldo}, J.~{Cohen-Tanugi},
  M.~{Ackermann}, T.~{Mizuno}, and F.~W. {Stecker}, {\it {On the Physics
  Connecting Cosmic Rays and Gamma Rays: Towards Determining the Interstellar
  Cosmic Ray Spectrum}},  {\em ArXiv e-prints} (Mar., 2013)
  [\href{http://arxiv.org/abs/1303.6482}{{\tt arXiv:1303.6482}}].

\bibitem{2013EPJWC..5202001O}
S.~{Ostapchenko}, {\it {QGSJET-II: physics, recent improvements, and results
  for air showers}},  in {\em European Physical Journal Web of Conferences},
  vol.~52 of {\em European Physical Journal Web of Conferences}, p.~2001, June,
  2013.

\bibitem{2002cra..book.....S}
R.~{Schlickeiser}, {\em {Cosmic ray astrophysics}}.
\newblock Cosmic ray astrophysics / Reinhard Schlickeiser, Astronomy and
  Astrophysics Library; Physics and Astronomy Online Library.~Berlin:
  Springer.~ISBN 3-540-66465-3, 2002, XV + 519 pp., 2002.

\bibitem{2015ApJ...802..114K}
S.~{Krakau} and R.~{Schlickeiser}, {\it {Pion Production Momentum Loss of
  Cosmic Ray Hadrons}},  {\em \apj} {\bf 802} (Apr., 2015) 114.

\bibitem{2012A&A...539A..88C}
B.~{Coste}, L.~{Derome}, D.~{Maurin}, and A.~{Putze}, {\it {Constraining
  Galactic cosmic-ray parameters with Z {$\le$} 2 nuclei}},  {\em \aap} {\bf
  539} (Mar., 2012) A88, [\href{http://arxiv.org/abs/1108.4349}{{\tt
  arXiv:1108.4349}}].

\bibitem{2011A&A...534A..54S}
A.~W. {Strong}, E.~{Orlando}, and T.~R. {Jaffe}, {\it {The interstellar
  cosmic-ray electron spectrum from synchrotron radiation and direct
  measurements}},  {\em \aap} {\bf 534} (Oct., 2011) A54,
  [\href{http://arxiv.org/abs/1108.4822}{{\tt arXiv:1108.4822}}].

\bibitem{2013MNRAS.436.2127O}
E.~{Orlando} and A.~{Strong}, {\it {Galactic synchrotron emission with cosmic
  ray propagation models}},  {\em \mnras} {\bf 436} (Dec., 2013) 2127--2142,
  [\href{http://arxiv.org/abs/1309.2947}{{\tt arXiv:1309.2947}}].

\bibitem{2015arXiv150606660P}
{Planck Collaboration}, {\it {Planck 2015 results. XXV. Diffuse low-frequency
  Galactic foregrounds}},  {\em ArXiv e-prints} (June, 2015)
  [\href{http://arxiv.org/abs/1506.0666}{{\tt arXiv:1506.0666}}].

\bibitem{2015APh....70...39K}
R.~{Kissmann}, M.~{Werner}, O.~{Reimer}, and A.~W. {Strong}, {\it {Propagation
  in 3D spiral-arm cosmic-ray source distribution models and secondary particle
  production using PICARD}},  {\em Astroparticle Physics} {\bf 70} (Oct., 2015)
  39--53, [\href{http://arxiv.org/abs/1504.0824}{{\tt arXiv:1504.0824}}].

\bibitem{2014APh....55...37K}
R.~{Kissmann}, {\it {PICARD: A novel code for the Galactic Cosmic Ray
  propagation problem}},  {\em Astroparticle Physics} {\bf 55} (Mar., 2014)
  37--50, [\href{http://arxiv.org/abs/1401.4035}{{\tt arXiv:1401.4035}}].

\bibitem{2012ApJ...757...14J}
R.~{Jansson} and G.~R. {Farrar}, {\it {A New Model of the Galactic Magnetic
  Field}},  {\em \apj} {\bf 757} (Sept., 2012) 14,
  [\href{http://arxiv.org/abs/1204.3662}{{\tt arXiv:1204.3662}}].

\bibitem{2012ApJ...761L..11J}
R.~{Jansson} and G.~R. {Farrar}, {\it {The Galactic Magnetic Field}},  {\em
  \apjl} {\bf 761} (Dec., 2012) L11,
  [\href{http://arxiv.org/abs/1210.7820}{{\tt arXiv:1210.7820}}].

\bibitem{2005ApJ...622..759G}
K.~M. {G{\'o}rski}, E.~{Hivon}, A.~J. {Banday}, B.~D. {Wandelt}, F.~K.
  {Hansen}, M.~{Reinecke}, and M.~{Bartelmann}, {\it {HEALPix: A Framework for
  High-Resolution Discretization and Fast Analysis of Data Distributed on the
  Sphere}},  {\em \apj} {\bf 622} (Apr., 2005) 759--771,
  [\href{http://arxiv.org/abs/astro-ph/}{{\tt astro-ph/}}].

\bibitem{2014A&A...569A..32M}
D.~{Maurin}, F.~{Melot}, and R.~{Taillet}, {\it {A database of charged cosmic
  rays}},  {\em \aap} {\bf 569} (Sept., 2014) A32,
  [\href{http://arxiv.org/abs/1302.5525}{{\tt arXiv:1302.5525}}].

\bibitem{2015arXiv150102003T}
{The Fermi-LAT Collaboration}, {\it {Fermi Large Area Telescope Third Source
  Catalog}},  {\em ArXiv e-prints} (Jan., 2015)
  [\href{http://arxiv.org/abs/1501.0200}{{\tt arXiv:1501.0200}}].

\bibitem{2015PhRvD..91h3012G}
D.~{Gaggero}, A.~{Urbano}, M.~{Valli}, and P.~{Ullio}, {\it {Gamma-ray sky
  points to radial gradients in cosmic-ray transport}},  {\em \prd} {\bf 91}
  (Apr., 2015) 083012, [\href{http://arxiv.org/abs/1411.7623}{{\tt
  arXiv:1411.7623}}].

\bibitem{2010A&A...516A..67M}
D.~{Maurin}, A.~{Putze}, and L.~{Derome}, {\it {Systematic uncertainties on the
  cosmic-ray transport parameters. Is it possible to reconcile B/C data with
  {$\delta$} = 1/3 or {$\delta$} = 1/2?}},  {\em \aap} {\bf 516} (June, 2010)
  A67, [\href{http://arxiv.org/abs/1001.0553}{{\tt arXiv:1001.0553}}].

\end{thebibliography}\endgroup



\end{document}